# 2.1 µm CW Raman Laser in GeO$_2$ Fiber

B. A. Cumberland, S. V. Popov and J. R. Taylor
Femtosecond Optics Group, Imperial College London, SW7 2AZ, United Kingdom

O. I. Medvedkov, S. A. Vasiliev, E. M. Dianov
Fiber Optics Research Center, General Physics Institute, Moscow 119333, Russian Federation

**Abstract**: We report on 33 % efficient generation of the first Stokes in a high concentration GeO$_2$ fiber Raman laser pumped by a 22 W Thulium doped fiber laser. An output power of 4.6 W at 2.105 µm is demonstrated.

Stimulated Raman scattering (SRS) has enabled the development of a wide variety of CW fiber lasers and amplifiers at wavelengths for which there are no rare-earth doped gain media available. This approached has also been applied to CW solid-state based systems.[1] Recently Raman lasers have been demonstrated in photonic crystal[2] and chalcogenide glass fibers[3] allowing extension of operational wavelengths of fiber lasers beyond the waveguiding loss and single mode propagation regions of conventional silica-based fibers. For generation of Raman fiber lasers with wavelengths beyond 2 µm the rapidly increasing losses of the bulk silica impose a limit on long wavelength operation of conventional silica fibers.

Alternative oxide-based glass forming hosts, such as high concentration GeO$_2$, were first suggested in 1975[4] as a material to produce low loss optical fibers. Shortly after a comparison of the Raman scattering cross-sections and Stokes intensity of vitreous bulk SiO$_2$, GeO$_2$, B$_2$O$_3$ and P$_2$O$_5$[5] indicated that the cross-section in GeO$_2$ is nearly an order of magnitude greater than that of SiO$_2$. It has also been shown that the intrinsic infrared absorption of GeO$_2$ glass is shifted to longer wavelengths as compared to silica glass because germanium atoms have a greater mass than silicon atoms, making GeO$_2$ a better candidate for Raman generation in the infrared. Doping of silica fibers with low concentrations of GeO$_2$ has been routinely used to enhance Raman gain. However manufacturing of high, above 20-40 mol.%, concentration GeO$_2$ core fibers has until recently faced problems.[6] These problems arise from the mismatch of the thermal expansion coefficients of GeO$_2$ and SiO$_2$ combined with the narrow temperature range between nonsintering and the evaporation of the GeO$_2$.

Manufacturing of single-mode fibers with core concentration of 51 to 97 mol.% GeO$_2$ was first reported in 2004.[6,7] Apart from lower losses above 2 µm these fibers possess a significantly enhanced nonlinearity as compared with silica fibers. The combination of these two advantages makes the highly doped GeO$_2$ fibers an excellent candidate for extending Stokes generation in optical fibers in the infrared. This potential has been demonstrated by using pump sources based on ytterbium and erbium doped fiber lasers.[6-8] As the Raman gain coefficient of such fibers is about an order of magnitude higher than that in silica fibers the increasing losses with wavelength can be compensated for by using much shorter lengths of fiber for Raman generation. Here we report on the efficient generation of a high power Raman source at 2.1 µm in a 75 mol.% $GeO_2$ fiber directly pumped with a single-mode thulium doped fiber laser. This approach allows for a high optical to optical and overall efficiency to be achieved in the wavelength region above 2 µm.

The experimental setup is shown in Fig. 1. A 22 W, 1.938 µm, ~1 nm linewidth, single-mode CW thulium (Tm) doped fiber laser (IPG Photonics) was employed as a pump. The free-space output of the Tm laser was bulk coupled into the $GeO_2$ fiber avoiding spurious back-reflection from the Raman laser cavity into the Tm laser. To reduce the thermal load in the coupling setup an optical chopper with a 25 % duty factor was used. The pump light was coupled into a short length of STF (standard telecom fiber) with an efficiency of up to 80 %. The STF was directly spliced to the 75 mol.% $GeO_2$ fiber using a mode-field matching technique on an arc-fusion splicer which resulted in regular splices losses of less than 0.5 dB. Two fiber Bragg gratings (FBG) at 2.105 µm corresponding to the first Raman shift of the pump wavelength, a high reflector (HR, R > 99 %) and output coupler (OC, R ~50 %), were written in the same $GeO_2$ fiber and formed the Raman laser cavity. Due to the high photosensitivity of the $GeO_2$ fiber the gratings were directly recorded with 244 nm laser radiation without hydrogen loading of the fiber.[9] The gratings' spectral properties were controlled during manufacturing by taking into account a calculated dispersion and measuring the second diffraction order. The fiber's core doping concentration was 75 mol.% of $GeO_2$ (25 mol.% $SiO_2$), it had a measured mode field diameter of 2.5 µm and a single mode cut off wavelength of ~1.4 µm. The measured loss in the 1.938 µm region was 21 dB/km increasing to 52-56 dB/km at 2.105 µm. This compares favorably to $SiO_2$ fiber, which has losses of ~16 and 110 dB/km at 1.94 and 2.11 µm respectively. A cut back experiment with five $GeO_2$ fiber cavity lengths of 10.3, 17.5, 26.3, 33.5 and 42.5 m was performed. Due to the high NA of the $GeO_2$ fiber a short length of STF (~5 cm) was spliced to the end of it to reduce the output NA for measurements. Spectral measurements were made using an automated Spex 500 spectrometer in combination with a PbS IR detector and lock-in amplifier.

The spontaneous Raman signals were first investigated by direct pumping of a 42.5 m long $GeO_2$ fiber at 20 W pump level resulting in signals at 2.113 µm and 2.322 µm (Fig. 2). Initially, the 42.5 m long cavity was built with HR and OC FBGs overlapping at 2.105 µm. This wavelength was slightly short of the peak spontaneous Raman gain but due to the broad $GeO_2$ Raman gain such detuning should not greatly affect the efficiency.[5] The cutbacks were made in order to maximize the output power at 2.105 µm. This was initially assessed by taking into account the fiber losses at 2.1 µm and the single-pass effective Raman interaction length, $(1-\exp(-\alpha L))\alpha^{-1}$, while ignoring the pump depletion. Here $\alpha$ is the fiber attenuation and L is the fiber's physical length. It can be seen from Fig. 3 that the resulting slope efficiency was identical for the 10.3 and 42.5 m cavities while in the 26.3 m long cavity the highest Raman generation efficiency was obtained. A maximum output power of 4.61 W (1.15 W average with the choppers duty factor of 25 %) with a FWHM bandwidth of 2.4 nm (Fig. 3 inset) was achieved in this cavity with a slope efficiency of 33 %, defined as the output signal power over the absorbed pump power. The lasing linewidth is related to the width of the HR and OC FBGs as the resolution of the spectrograph was in the sub-nm range.

Fig. 4 illustrates the residual pump power dynamics versus the input pump power level. For the 26, 33 and 42 m cavities the residual pump power remains largely saturated within the 1.5 to 2.0 W region. Optimization of the gain-loss balance by taking into account the pump depletion and by changing the OC FBG reflectivity should allow higher efficiencies. Further improvements to the current cavity geometry may be possible by utilizing the residual pump in the cavity through the addition of a blocking FBG at the pump wavelength after the OC grating.

We estimated the Raman gain, $g_R$, at 2.1 µm by taking into account the experimentally measured cavity losses and threshold pump powers as 3.7±0.4/W/km. The error in this case is associated with the uncertainty in reflectivity value of the OC FBG. The estimated value of $g_R$ is over an order of magnitude higher than the peak Raman gain of STF (~0.5/W/km at 2.11 µm).

The much weaker second order Stokes at 2.322 µm, observed from spontaneous Raman generation (Fig. 2), offers an opportunity to produce a Raman laser at this wavelength. However an estimated 5-fold increase in transmission losses and a reduction in the Raman gain proportional to the inverse wavelength will require the use of short linear cavities, in similar or reduced concentration $GeO_2$ fibers, in order to achieve Raman generation at this wavelength.

In conclusion, we have demonstrated a 4.6 W CW generation around 2.1 µm in a linear Raman laser cavity based on a 75 mol.% $GeO_2$ fiber. The high concentration germanium fiber provided significantly enhanced Raman gain and lower optical losses above 2 µm compared with standard silica fibers. By employing direct pumping with a 1.938 µm thulium fiber laser a slope efficiency of 33 % at 2.1 µm was achieved. The detection of a weak spontaneous Raman signal at 2.3 µm opens up the prospect of a Raman laser at this wavelength although the considerable increase in fiber losses at this wavelength may limit the output powers.

Note that since the submission of this paper utilization of a HR FBG at 1.938 µm has resulted in a maximum output power of 6.8W (1.7W average) at 2.1 µm from a 6.7 m cavity with a slope efficiency of 58%. The calculated Raman gain for this cavity is 61±8/W/km, which is much closer to the expected value of 70/W/km.

The authors acknowledge the following sources of support: B. A. Cumberland is funded by an EPSRC studentship. S. V. Popov is a Royal Society Industrial Fellow. J. R. Taylor is a Royal Society Wolfson Research Merit Award holder.

arxiv.org:physics/0703083 v3 21 June 2007

**Figures**

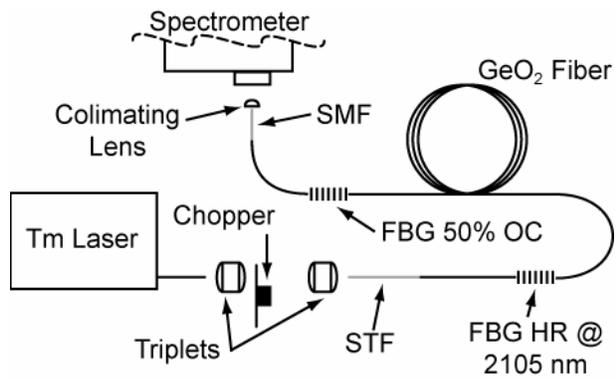

Fig. 1. Schematic of the experimental setup.

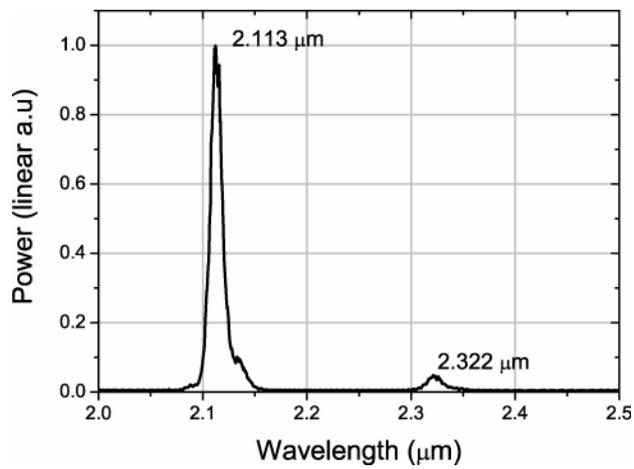

Fig. 2. Spontaneous Raman signal at 2.113 µm and 2.322 µm from 42 m of $GeO_2$ fiber.

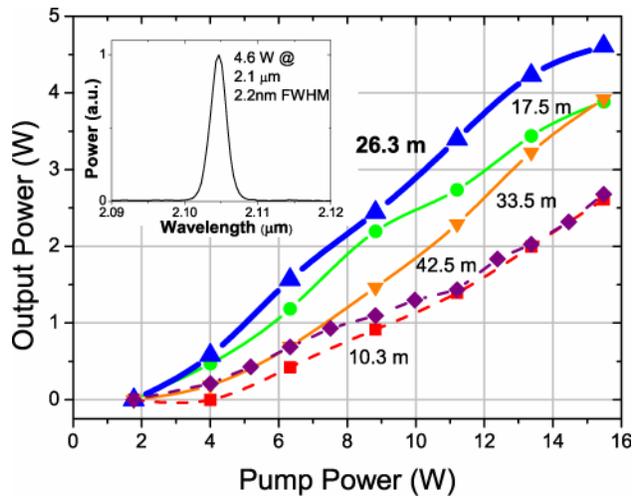

Fig. 3. (Color online). 2.105 µm Raman output power *vs* coupled input pump power for five cavity lengths: 10.3 m (squares), 17.5 m (circles), 26.3 m (triangles point up), 33.5 m (triangles point down), 42.5 m (diamonds). Inset: spectral output from the 26.3 m cavity with 4.6 W of output power.

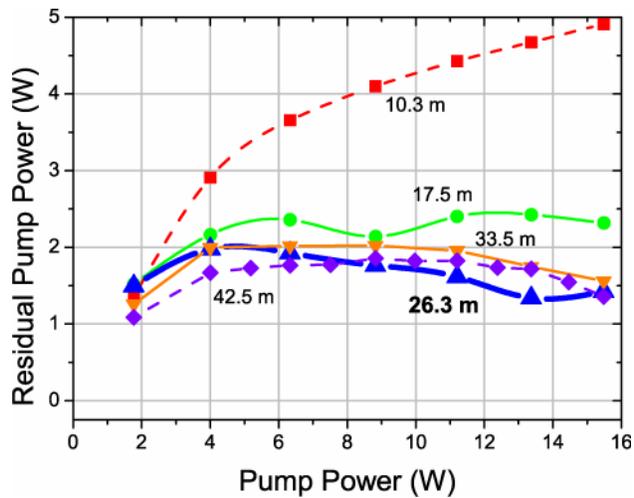

Fig. 4. (Color online). Residual pump power at 1.938 µm *vs* coupled input pump power for five cavity lengths: 10.3 m (squares), 17.5 m (circles), 26.3 m (triangles point up), 33.5 m (triangles point down), 42.5 m (diamonds).